\documentclass[preprint,12pt]{elsarticle}
\usepackage{amssymb}
\usepackage{graphicx}
\usepackage{caption}
\usepackage{natbib}
\usepackage{dcolumn}
\usepackage{longtable}
\usepackage{subfigure}
\usepackage{rotating}
\usepackage{lscape}
\usepackage{mathrsfs}
\usepackage{amsmath}

\journal{Materials Horizons}

\begin{document}
\begin{frontmatter}



\title{Two-dimensional ferromagnetic semiconductor Cr$_{2}$XP: First-principles calculations and Monte Carlo simulations}
\author{Xiao-Ping Wei$^{a,b}$}
\ead{weixp2008@lzjtu.edu.cn}
\author{Lan-Lan Du$^{a}$}
\author{Jiang-Liu Meng$^{a}$}
\author{Xiaoma Tao$^{c}$}
\address{$^{a}$The School of Mathematics and Physics, Lanzhou Jiaotong University, Lanzhou 730070, P. R. China.}
\address{$^{b}$Lanzhou Center for Theoretical Physics, Key Laboratory of Theoretical Physics of Gansu Province, Lanzhou University, Lanzhou, Gansu 730000, P. R. China}
\address{$^{c}$College of Physical Science and Technology, Guangxi University, Nanning 530004, P. R. China.}

\begin{abstract}
According to the Mermin Wagner theorem, two-dimensional material is difficult to have the Curie temperature above room temperature. By using the method of band engineering, we design a promising two-dimensional ferromagnetic semiconductor Cr$_{2}$XP (X=P, As, Sb) with large magnetization, high Curie temperature and sizable band gap. The formation of gap is discussed in terms of the hybridizations, occupation and distribution of electronic states and charge transfer. Large magnetic moments about 6.16$\sim$6.37$\mu_{B}$ origin from the occupation of Cr-$d$ electrons in crystal field. Competition and cooperation between $\textit{d-d}$ (Cr-$\textit{d}$$\sim$Cr-$\textit{d}$) and $\textit{d-p-d}$ (Cr-$\textit{d}$$\sim$X-$\textit{p}$$\sim$Cr-$\textit{d}$) exchange interactions lead to the emergence of ferromagnetic ordering phase. Furthermore, Curie temperatures, approaching to 269 K, 332 K and 400 K for Cr$_{2}$P$_{2}$, Cr$_{2}$AsP and Cr$_{2}$SbP, are estimated by employing Monte Carlo simulation based on the Heisenberg model. Magnetic anisotropy energy of Cr$_{2}$XP is determined by calculating the total energy dependence on the angle along different directions, and the origin is also discussed by the second-order perturbation theory. In addition, the Cr$_{2}$XP possesses excellent thermodynamical, dynamical and mechanical stabilities, and can overcome their own gravity to keep their planar structure without the support of substrate. These above-mentioned advantages will offer some valuable hints for two-dimensional ferromagnetic semiconductor Cr$_{2}$XP in spintronic devices.
\end{abstract}


\end{frontmatter}


\section{Introduction}

Two-dimensional (2D) materials are widely used in energy storage and conversion, thermal management and spintronic devices \cite{Y.D,M.X,H.Z} due to their excellent mechanical strength, flexibility, optical transparency and a large specific surface area etc. In particular, 2D spintronic materials, such as topological insulators, magnetic metals and magnetic semiconductors, offer opportunities for the development of next-generation information technology because of their strong non-volatility, lower power consumption, good stability, fast information processing and high integration \cite{E.C,A.F,S.A}. \par
Among 2D materials, ferromagnetic semiconductors have been found to possess both ferromagnetism and semiconductor characteristics. Furthermore, the continuously changing magnetic polarization direction of ferromagnetic materials makes them appropriate for applications in magnetic storage or spin valves. Currently, 2D ferromagnetic semiconductor materials that have been reported include GdI$_{2}$ \cite{W.B}, CrI$_{3}$ \cite{B.H.X}, CrX$_{3}$ (X=F, Cl, Br, I) \cite{W.B.Z}, CrWI$_{6}$ \cite{C.H}, CrSX (X=Cl, Br, I) \cite{K.Y}, CrSnTe$_{3}$ \cite{H.L.Z}, CrOF \cite{C.X},  MnNCl and CrCX (X=Cl, Br, I; C=S, Se, Te) \cite{C.W}. However, their application in spintronic devices is limited by their low Curie temperature, small magnetization and magnetic anisotropy energy. In 2017, B. Huang et al. discovered that monolayer CrI$_{3}$ is a two-dimensional ferromagnet with in-plane spin orientation. It exhibits a Curie temperature of 45 K, slightly lower than the bulk crystal's 61 K, consistent with weak interlayer coupling. This finding, through the use of magnetic anisotropy, breaks the prohibition of magnetic ordering in the 2D isotropic Heisenberg model at finite temperatures, opening up possibilities in 2D ferromagnetic semiconductor materials \cite{B.H.X}. Based on this foundation, strong ferromagnetism and semiconducting behavior have been observed in the 2D monolayer CrX$_{3}$ (X=F, Cl, Br, I). Research has shown that CrX$_{3}$ not only exhibits a large magnetic moment of 3$\mu_{B}$ and significant magnetic anisotropy, but also reveals competition between nearly 90$^{\circ}$ FM super-exchange and direct anti-ferromagnetic (AFM) exchange, resulting in FM nearest-neighbor exchange interactions. However, a drawback is that their Curie temperatures are below room temperature, severely limiting their applications in spintronic devices \cite{W.B.Z}. Consequently, the search for high Curie temperature 2D ferromagnetic semiconductor material has become a major area of interest for researchers. Subsequently, the MnNCl and CrCX (X=Cl, Br and I; C=S, Se and Te) were discovered as ferromagnetic semiconductors, Monte Carlo simulations based on the Heisenberg model predicted their Curie temperatures ranging beyond 100 K to nearly 500 K \cite{C.W}. Furthermore, these materials exhibited wide band gaps and significant magnetic anisotropy. However, the mechanisms responsible for the ferromagnetic phase and magnetic anisotropy were not explained. In contrast, CrWI$_{6}$ is elucidated the mechanism of ferromagnetic coupling between magnetic atoms, serving as a general mechanism to significantly enhance the ferromagnetic coupling of 2D semiconductor materials without the introduction of charge carriers \cite{C.H}. Additionally, it had a magnetic anisotropy of 5.4 meV, making it a candidate material for spintronics, because 2D ferromagnetic materials with large magnetic anisotropy have tremendous potential in realising low critical switch currents, high thermal stability, and high-density non-volatile storage in magnetic random access memory. For instance, the recently discovered 2D ferromagnetic semiconductor materials CrOF and CrOCl exhibit significant magnetic anisotropy and a large total magnetic moment of $6\mu_{B}$. However, their Curie temperatures are 277 K and 120 K, which are evident below room temperature, limiting their application in the spintronic field \cite{C.X}. 2D transition metal trichalcogenide materials with the chemical formula Cr$_{2}$X$_{2}$Se$_{6}$ (X=Al, Ga, In) was predicted to be semiconductors, and the size of band gap is decreasing as X changes, the band gaps of Cr$_{2}$Al$_{2}$Se$_{6}$, Cr$_{2}$Ga$_{2}$Se$_{6}$ and Cr$_{2}$In$_{2}$Se$_{6}$ are 1.29 eV, 0.91 eV and 0.86 eV, respectively. Single-layer Cr$_{2}$X$_{2}$Se$_{6}$ exhibits ferromagnetic ordering and out-of-plane magnetic anisotropy (MAE) \cite{M.M}. Consequently, the development of 2D ferromagnetic materials with large magnetization, sizeable magnetic anisotropy and high Curie temperature is of great importance for the development of devices in post-Moore era devices.\par

In the current research, we systematically investigated the electronic structure, magnetic properties, and Curie temperature of the Cr$_{2}$XP system. Our research revealed that Cr$_{2}$XP exhibits high magnetization, significant magnetic anisotropy, and a Curie temperature above room temperature. These outstanding properties suggest that the Cr$_{2}$XP system has the potential to become a candidate material for spintronics. \par

\section{Calculation details}
Electronic calculations are carried out by using spin-polarized density functional theory within the Vienna \textit{ab-initio} Simulation Package (VASP) along with the projector augmented-wave (PAW) potentials \cite{G.K,G.K.D}. The Perdew-Burke-Ernzerhof (PBE) within the generalized gradient approximation (GGA) is used to consider the exchange correlation effects \cite{L.A.C}. During the whole process of calculations, the plane wave truncation energy of 500 eV was set. The geometry structure was optimized by using the conjugate gradient method and the Methfessel-Paxton smearing method, and the criterion of force on each atom was -0.02 eV/{\AA} in the fully optimization, and the convergence of electron iteration was set to 10$^{-6}$ eV/atom. The Brillouin zone was sampled by using 21$\times$21$\times$1 Monkhorst-Pack k-points. A 15 {\AA} vacuum was applied to avoid the interactions between layers. The phonon spectrum was calculated by using the PHONOPY program \cite{T.A}. \par

\section{Results and discussion}
\subsection{Crystal structure and stability}

Figure 1 depicts the crystal structure of 2D Cr$_{2}$XP (X=P, As, Sb) with P4/nmm space group (No. 129) \cite{A.N.M}. Subsequently, we construct the different magnetic configurations in Figure 2(a), and listed their total energy by calculations in Table 1. It is found that 2D Cr$_{2}$XP is ferromagnetic ground state. In Table 2, we show the optimized lattice constants of 2D Cr$_{2}$XP (X=P, As, Sb) based on the ferromagnetic ground state, they are 4.17 {\AA}, 4.23 {\AA} and 4.29 {\AA}, respectively. \par

To estimate the thermodynamic stability of 2D Cr$_{2}$XP, we calculate their formation energy $E_{\textit{f}}$ in the following formula:
\begin{equation}
E_{f}=\frac{1}{N_{atom}}[E_{Cr_{2}XY}-(2E_{Cr}+E_{X}+E_{Y}) ]
\end{equation}
where $E_{Cr_{2}XY}$ is the total energy per unit cell, $N_{atom}$ is the total atomic number in the formula unit, $E_{Cr}$, $E_{X}$ and $E_{Y}$ are their atomic chemical potentials. It can be seen from Table 2 that the formation energies of Cr$_{2}$XP are less than zero, implying they are easier to be prepared in experiment and meet the thermodynamic stability.\par

In order to evaluate the dynamical stability, we calculate the phonon spectrum of Cr$_{2}$XP, which describes the variation of the total energy relative to the atomic displacement, and the total energy can be shown by the shifts $D_{\textbf{R}\sigma}$ in the conditions of harmonic approximation \cite{D.O.G}:
\begin{equation}
 E=E_{0}+\frac{1}{2}\sum_{\bf R,\sigma}\sum_{\bf R',\sigma'}D_{\bf R\sigma}\Phi^{\sigma\sigma'}_{\bf R\bf R'} D_{\bf R'\sigma'}
\end{equation}
where $\textbf{R}$ is the position, $\sigma$ is the Cartesian index, $\Phi^{\sigma\sigma'}_{\textbf{RR}'}$ is the interatomic force constant matrix. Dynamic matrix $D(\bf\textbf{q})$ determines the stability of the crystal, and can be acquired by performing a Fourier transform of $\Phi(\textbf{R})$ in the following:
\begin{equation}
D{\bf (q)}=\frac{1}{M}\sum_{R}\Phi({\bf R})e^{-i\bf qR}
\end{equation}
we notice from Figure 3 that Cr$_{2}$XP does not have imaginary frequency, thus it satisfies dynamic stability.\par

Subsequently, we assess the mechanical stability by calculating the elastic constants. The calculated elastic constants are rescaled in the following equation:
\begin{equation}
C_{ij}=C_{ij,cell}\cdot c
\end{equation}
where $C_{ij,cell}$ is calculated elastic constants with the vacuum and $c$ is the lattice parameter in the $z$ direction \cite{X.H,X.P}. The mechanical stability can be estimated by the Born-Huang criteria \cite{M.B}:
\begin{equation}
C_{11}> 0,
C_{12}> 0,
C_{66}> 0,
C_{11}C_{22} - C_{12}^{2} > 0,
\end{equation}
It can be seen from Table 2 that the elastic constants of 2D monolayer Cr$_{2}$XP satisfy the Born-Huang criteria, therefore, they possess the mechanical stability.\par

Finally, we estimate the typical plane deformation caused by gravity in $h/\textit{l}$ $\approx$ ($\sigma$g$\textit{l}$/$Y_{s}^{2D}$)$^{\frac{1}{3}}$,  where $\sigma$ is the surface density, the calculated surface density is 0.962$\times$10$^{-6}$, 1.161$\times$10$^{-6}$ and 1.319$\times$10$^{-6}$ for Cr$_{2}$P$_{2}$, Cr$_{2}$AsP and Cr$_{2}$SbP, the $\textit{l}$ is evaluated size, if we assume $\textit{l}$=100$\mu$m, and then $h$/$\textit{l}$ is 3.23$\times$10$^{-4}$, 3.53$\times$10$^{-4}$ and 3.78$\times$10$^{-4}$ for monolayer Cr$_{2}$P$_{2}$, Cr$_{2}$AsP and Cr$_{2}$SbP, respectively. Obviously, these values have the same order of magnitude as that of graphene \cite{T.J},  indicating that the Cr$_{2}$XP can overcome their own gravity to keep their planar structure without the support of substrate.\par

\subsection{Elastic properties}
Young's modulus ($Y^{2D}_{s}$), shear modulus($G_{2D}$) and Poisson's ratio ($\upsilon$) are crucial parameters for evaluating the mechanical properties of 2D materials. $Y^{2D}_{s}$ denotes the resistance of solid material to deformation. $\upsilon$ is the ratio of the lateral strain to the axial strain when a material is subjected to uniaxial tension or compression, it is an elastic constant that reflects a material's transverse deformation. $G_{2D}$ is the ratio of shear stress to strain, representing the material's ability to resist shear strain. Based on the calculated elastic constants, $Y^{2D}_{s}$, $G_{2D}$ and $\upsilon$ can be obtained as follows \cite{X.P.W}:
\begin{equation}
	Y_{s}^{2D}=\frac{C_{11}^2-C_{12}^2}{C_{11}}
\end{equation}
\begin{equation}
	G_{2D}=C_{66}
\end{equation}
\begin{equation}
	\upsilon=\frac{C_{12}}{C_{11}}
\end{equation}
It can be seen from Table 3 that the maximum Young's modulus of 2D Cr$_{2}$XP is significantly lower than that of MoS$_{2}$ (129 N/m) \cite{S.J.C}, graphene (340 N/m) \cite{R.C.A}, and h-BN (255.8 N/m) \cite{K.H.M}, implying that 2D Cr$_{2}$XP exhibits their lower in-plane stiffness. The Poisson's ratio of 2D Cr$_{2}$XP is significantly higher than those of MoS$_{2}$ (0.16), graphene ($\sim$0.1) and h-BN (0.15$\sim$0.21), indicating a stronger transverse deformation. The shear modulus of 2D Cr$_{2}$XP is lower than that of MoS$_{2}$ (150.5 N/m), graphene (46.3 N/m), and h-BN (117.85 N/m), indicating their lower ability to resist shear strain. In addition, we used ELATE software \cite{R.G} to obtain 2D elastic modulus diagrams of Cr$_{2}$XP in Figure 4. It is evident that the Young's modulus, shear modulus and Poisson's ratio of 2D Cr$_{2}$XP display an anisotropic behaviour. \par

\subsection{Electronic properties}
In order to obtain more accurate electronic properties, we additionally use the Heyd-Scuseria-Ernzerhof hybrid functional method (HSE06) \cite{S.L} to calculate the band structures and resolved orbital density of states (DOS) of Cr$_{2}$XP. It can be observed in Figure 5 that the spin-up and spin-down bands do not overlap at the Fermi level, indicating that Cr$_{2}$XP is a ferromagnetic semiconductor. In particular, the spin-down gap is bigger than that of spin-up because of stronger splitting of Cr-3$d$ orbitals. Also, we notice that the Cr-3$d$ orbitals offer a main contribution to the spin-up and spin-down total density of states above the Fermi level. The $\textit{p}$ states primarily contribute to the spin-down states below the Fermi level, and more contribution for spin-up states come mainly from the Cr-3$d$ states below the Fermi level. More importantly, a hybridization between $d$ and $p$ orbitals enhances the stability of these systems.\par

Next, we focus on the origin of band gap by calculating the local density of states (LDOS) of Cr$_{2}$XP (X=P, As, Sb). In Figure 6, we give the LDOS of Cr-3$d$ and P-, As- and Sb-$p$ states. It can be observed that the contribution of Cr-3$d$ orbitals is larger than that of the P-$\textit{p}$, As-$\textit{p}$, and Sb-$\textit{p}$ orbitals. Specifically, we take Cr$_{2}$P$_{2}$ as an example to discuss the origin of the gap. We notice that the $d$ orbtial is split into double-degenerate $e_{1}$ ($d_{xz}$ and $d_{yz}$) states and singlet states $\textit{a}$$_{1}$ ($d_{z^{2}}$), $b_{1}$ ($d_{x^{2}-y^{2}}$) and $b_{2}$ ($d_{xy}$). For the P-$\textit{p}$ state, it is split into double-degenerate $\textit{a}$$_{2}$ ($\textit{p}$$_{x}$ and $\textit{p}$$_{y}$) and singlet state $e_{3}$ ($\textit{p}$$_{z}$). The unoccupied P-$p$ state provides a main contribution in the range of -4$\sim$-1 eV in spin-down channel, and the Cr-3$d$ states offer a main contribution to spin-up total DOS between -4$\sim$-1 eV, also play a leading role in 2$\sim$8 eV of spin-down channel. Thereby, we can conclude that the exchange splitting of Cr-3$d$ play a role for the formation of gap in 2D Cr$_{2}$P$_{2}$. Accordingly, similar conclusions can also be obtained in Cr$_{2}$AsP and Cr$_{2}$SbP. Furthermore, we offer the differential charge density in Figure 7 to consider whether the electron transfer has an influence on the formation of the gap, where yellow and blue colors represent the accumulation and depletion of charge, respectively. It can be seen that the Cr atom loses electrons, and the P, As and Sb atoms gain electrons, this is also consistent with the Bader charge in Table 4. Finally, we believe that the exchange splitting of $d$ and $p$ electronic states and atomic charge transfer together lead to the formation of gap. Among them, the exchange splitting of $d$ states plays a leading role in the origin of the gap. In addition, we also show the occupation and distribution of $d$ states in the crystal field. It is found in Figure 8 that the Cr-3$d$ states are split into double-degenerate $e_{1}$ ($d_{xz}$ and $d_{yz}$) states and singlet states $\textit{a}$$_{1}$ ($d_{z^{2}}$), $b_{1}$ ($d_{x^{2}-y^{2}}$) and $b_{2}$ ($d_{xy}$) in 2D square lattice field, wherein, the singlet state $a{_1}$ and two degenerate $e{_1}$ states are located below the Fermi level, and singlet states $b{_1}$ and $b{_2}$ are distributed above the Fermi level. Particularly, the $a{_1}$ and $b{_2}$ states determine the size of gap. \par

In order to study the bonding behavior in Cr$_{2}$XP system, we calculate the electron localization function (ELF), which can well describe electron localization in molecules and solids. The ELF is defined as \cite{S.A.J,S.K}
\begin{equation}
ELF(r) =\frac{1}{\left[1+D(r)/D_{h}(r)\right]^{2}}
\end{equation}
Usually, large ELF values correspond to covalent bonds and inner shell or lone pair electrons, whereas electron deficiency and ionic or metallic bonds are shown by smaller values \cite{K.K}. Here, ELF maps and ELF line profiles connecting bonding atoms are plotted for Cr$_{2}$XP system, and highlighted in Figure 9.
The ELF values of Cr atoms are 0.0 at both ends of the distance, suggesting its electron deficiency in these systems, and the ELF values for P, As and Sb atoms are not equal to 0.0, indicating they gain the electrons in systems. Accordingly, the electrons are obviously transferred from Cr atoms to others, which is also confirmed by the Bader charge in Table 4. For the Cr$_{2}$P$_{2}$, there is a medium ELF value of $\sim$0.55 in the intermediate region between Cr and P atoms, indicating a weak covalent bonding. The ELF value $\sim$0.3 between Cr and P atoms shows an ionic bonding behavior.
The ELF line profiles between two neighboring P atoms in Figure 8(d) show that there are obvious phosphorophilic interactions as shown by the dashed box. In the Cr$_{2}$AsP and Cr$_{2}$AsSb, the type of bonding between Cr atoms is consistent with the that of Cr$_{2}$P$_{2}$. Furthermore, the bonding of Cr and As atoms as well as Cr and Sb atoms is the same as that of the Cr and P atoms. Additionnally, the type of bonding between P atoms is also the same as that of As and P atoms as well as Sb and P atoms. In particular, the neighboring As atoms and Sb atoms in Figure 9(e) and (f) show obvious asophilicity and antimony affinity because of their increased distance. \par

\subsection{Magnetic properties}
To discuss the magnetic properties of Cr$_{2}$XP (X=P, As, Sb), we firstly picture the spin charge density for Cr$_{2}$XP in Figure 10. It can be seen that the spin magnetic moment is mainly contributed by the Cr atom, which is well consistent with the calculated atomic magnetic moment in Table 5. The magnetic moments of X and P atoms are rather small because of their electrons lower polarization. We can see from Table 5 that the total magnetic moments of Cr$_{2}$XP (X=P, As, Sb) are 6.165, 6.237 and 6.372$\mu_{B}$, and the magnetic moments of Cr atoms are 3.439, 3.511 and 3.603$\mu_{B}$, playing a main contribution to the total one. Furthermore, the magnetic moments of X and P atoms about $\sim$-0.4$\mu_{B}$ almost have no change, thus Cr$_{2}$XP is ferromagnet. \par

In the following, we discuss the formation of Cr atomic moment by the Bader charge transfer and the occupation of Cr-$d$ electron in crystal field. Here, we omit the X and P atoms because of their rather small magnetic moments. As is known to all, the outermost electrons in the Cr-$d$ orbital are five, losing an electron in Cr$_{2}$XP (X=P, As, Sb) results in the Cr$^{+1}$ (see Table 4). According to the splitting and occupation Cr-$d$ in 2D square lattice field, the $b_{1}$ ($d_{x^{2}-y^{2}}$) and $b_{2}$ ($d_{xy}$) states are distributed above the Fermi level, and the $a_{1}$($d_{z^{2}}$) and $e_{1}$ ($d_{xz}$ and $d_{yz}$) states are localized below the Fermi level. The remaining four electrons are distributed separately in $b_{2}$, $a_{1}$ and $e_{1}$ states. Among them, the $e_{1}$ state accommodates two electrons due to its two degenerate states, thus the Cr atomic magnetic moment is close to 4$\mu_{B}$.\par

\subsection{Magnetic anisotropy and Curie temperature}
A long-range magnetic order in 2D system requires a large magnetic anisotropy energy (MAE). The MAE is defined as MAE=E$_{\|}$-E$_{\bot}$, where E$_{\|}$ and E$_{\bot}$ represent the total energy in-plane and out-of-plane, respectively, after considering the spin-orbit coupling (SOC) interaction. Initially, we calculate the MAEs dependent on the angles in the $xy$, $xz$ and $yz$ planes as shown in Figure 11, it can be seen that Cr$_{2}$P$_{2}$ is an out-of-plane magnetic anisotropy material, whereas the Cr$_{2}$AsP and Cr$_{2}$SbP are in-plane magnetic anisotropy materials, the results are consistent with the calculated data in Table 6. The MAEs of Cr$_{2}$P$_{2}$, Cr$_{2}$AsP and Cr$_{2}$SbP are 60.97, -258.11 and -1904.97 $\mu$eV, respectively, ensuring the thermal stability of Cr$_{2}$XP as a magnetic storage material, and making them better suitable for achieving long-range magnetic order in 2D magnetic materials.\par

Next, we qualitatively analyze the origin of MAE by using the second-order perturbation theory \cite{G.X.W}:
\begin{equation}
\begin{aligned}\label{10}
  MAE&=\sum_{\sigma,\sigma'}(2\delta_{\sigma\sigma'}-1)\xi^{2}\times\sum_{\textit{o}^{\sigma},u^{\sigma'}}\frac{|\langle\textit{o}^{\sigma}|L_{\textit{z}}|u^{\sigma'}\rangle|^{2}-|\langle\textit{o}^{\sigma}|L_{\textit{x}}|u^{\sigma'}\rangle|^{2}}{E^{\sigma'}_{u}-E^{\sigma}_{\textit{o}}}
\end{aligned}
\end{equation}
where $\xi$ denotes the SOC constant, $E^{\sigma'}_{u}$ and $E^{\sigma}_{\textit{o}}$ are the energy levels of unoccupied states with spin $\sigma' $ ($\langle$u$^{\sigma'}|$) and occupied states with spin $\sigma$ ($\langle$$\textit{o}$$^{\sigma}|$), respectively. The $|\langle\textit{o}^{\sigma}|L_{\textit{z}}|u^{\sigma'}\rangle|^{2}-|\langle\textit{o}^{\sigma}|L_{\textit{x}}|u^{\sigma'}\rangle|^{2}$ was square the difference between angular momentum matrix element. It is the fact that the magnetic moment of Cr$_{2}$XP mainly originates from the Cr-$d$ orbitals, we specifically analyzed the origin of MAE for Cr$_{2}$XP by the PDOS and $d$ orbital contribution of MAE as displayed in Figure 12.\par

The MAE is contributed by the spin-up, spin-down and their mixed electronic states. In the 2D system, the non-zero matrix elements of the $d$ angular momentum operators $L_{x}$, $L_{y}$ and $L_{z}$ are as follows \cite{D.W.R}: $\langle d_{z^{2}}|L_{x}|d_{yz}\rangle$=$\sqrt{3}$, $\langle d_{xy}|L_{x}|d_{xz}\rangle$=$1$, $\langle d_{x^{2}-y^{2}}|L_{x}|d_{yz}\rangle$=$1$; $\langle d_{z^{2}}|L_{y}|d_{xz}\rangle$=$\sqrt{3}$, $\langle d_{xy}|L_{y}|d_{yz}\rangle$=$1$, $\langle d_{x^{2}-y^{2}}|L_{y}|d_{xz}\rangle$=$1$; $\langle d_{xz}|L_{z}|d_{yz}\rangle$=$1$, $\langle d_{x^{2}-y^{2}}|L_{z}|d_{xy}\rangle$=$2$. Then we take Cr$_{2}$P$_{2}$ as an example to discuss the origin of MAE, we can notice from the Figure 12(a) that there is a strong coupling between the spin-down occupied state $d_{x^{2}-y^{2}}$ and the spin-up unoccupied state $d_{xy}$, as well as between the spin-up occupied state $d_{z^{2}}$ and the spin-up unoccupied states $d_{yz}$ around the Fermi level. Specifically, the matrix elements $|\langle d_{xy}|d_{xz}\rangle|^{2}$, $|\langle d_{x^2-y^2}|d_{yz}\rangle|^{2}$ and $|\langle d_{z^{2}}|d_{yz}\rangle|^{2}$ offer a positive value for MAE. However, the $|\langle d_{xz}|d_{yz}\rangle|^{2}$  and $|\langle d_{x^{2}-y^{2}}|d_{xy}\rangle|^{2}$ have negative values, which decrease the MAE. The positive MAE value is bigger than that of the negative as shown in Figure 12(d), thus the Cr$_{2}$P$_{2}$ has an easy magnetization axis along the $z$-axis direction, and indicating a perpendicular magnetic anisotropy. For the Cr$_{2}$AsP and Cr$_{2}$SbP, although the matrix elements show the same positive and negative symbols as seen in Cr$_{2}$P$_{2}$, the negative MAE from $|\langle d_{x^{2}-y^{2}}|d_{xy}\rangle|^{2}$ plays a significant role, thus they exhibit the in-plane easy magnetization axis as displayed in Figure 12(e) and (f).\par

In order to obtain the Curie temperature of Cr$_{2}$XP, we first construct the spin Hamiltonian of the system as follows:
\begin{equation}
  H=-\sum_{{\textit{i},\textit{j}}}J_{1}S_{\textit{i}}S_{\textit{j}}-\sum_{{\textit{i},\textit{k}}}J_{2}S_{\textit{i}}S_{\textit{k}}-AS^{\textit{z}}_{\textit{i}}S^{\textit{z}}_{\textit{i}}
\end{equation}
where $\textit{S}$ represents the spin of the Cr atom, which equals to 3/2, and $J_{1}$ and $J_{2}$ are the exchange coupling parameters between the nearest and next-nearest neighbors (see Figure 2(b)), they are determined by calculating the total energy difference between the FM and AFM configurations in the following equation:
\begin{equation}
  E(FM)=E_{0}-16J_{1}S^{2}-16J_{2}S^{2}-AS^{2}
\end{equation}
\begin{equation}
  E(AFM1)=E_{0}+16J_{1}S^{2}-16J_{2}S^{2}-AS^{2}
\end{equation}
\begin{equation}
  E(AFM3)=E_{0}+16J_{2}S^{2}-AS^{2}
\end{equation}
where $E_{0}$ is the total energy without spin polarization. The $J_{1}$ and $J_{2}$ values for Cr$_{2}$P$_{2}$, Cr$_{2}$AsP and Cr$_{2}$SbP are 42.4 and 21.9 meV, 39.1 and 21.3 meV, 29.6 and 18.2 meV, respectively. Based on the calculated exchange coupling parameters, we simulate the magnetization and specific heat capacity dependence on temperature by using the Monte Carlo method as depicted in Figure 13. The calculated Curie temperatures of Cr$_{2}$P$_{2}$, Cr$_{2}$AsP and Cr$_{2}$SbP are 269 K, 332 K, and 400 K, respectively. Notably, the Curie temperatures of Cr$_{2}$AsP and Cr$_{2}$SbP are evidently higher than room temperature, which holds significant implications for their application in spintronic devices. Finally, we offer the MAE, total magnetic moment, Curie temperature and band gap for other reported 2D magnetic semiconductor for comparison in Figure 14. The results show that Cr$_{2}$XP exhibits significantly high Curie temperature and large magnetization, making it a potential candidate for spintronic applications.
\section{Conclusion}
We systematically study the stability, electronic and magnetic properties as well as magnetic anisotropy and Curie temperature for 2D Cr$_{2}$XP (X=P, As, Sb) by using first-principles calculations and Monte Carlo simulation. Results show that Cr$_{2}$XP meets the thermodynamic, dynamical and mechanical stabilities, and can overcome their own gravity to keep their planar structure
without the support of a substrate. Electronic calculations predict Cr$_{2}$XP (X=P, As, Sb) to be ferromagnetic semiconductor with gap 0.4987, 0.4922 and 0.0956 eV. The band gap is chiefly ascribed to the exchange splitting of Cr-$d$ state and charge transfer. The total spin magnetic moments of Cr$_{2}$XP (X=P, As, Sb) are 6.165, 6.237 and 6.372$\mu{_B}$, and mainly contributed by that of the Cr atom, which originates from the occupation of electrons in $b{_2}$, $a{_1}$ and $e{_1}$ states.
The competition between $d$-$d$ direct exchange and super-exchange $d$-$p$-$d$ leads to the emergence of ferromagnetic ordered phase for Cr$_{2}$XP.
The MAEs of Cr$_{2}$XP (X=P, As, Sb) are 60.97, -258.11 and -1904.97$\mu$eV, and the Cr$_{2}$P$_{2}$ is the easy magnetization out of the plane, while the Cr$_{2}$AsP and Cr$_{2}$SbP have an in-plane easy magnetization axis. The Cr-$d$ orbital provides a leading contribution to the MAEs. Finally, the Curie temperatures of Cr$_{2}$XP (X=As, Sb) are evidently higher than room temperature up to 332 K and 400 K. Our research may offer some valuable hints for the application of Cr$_{2}$XP in spintronics devices.\par
\section{Acknowledgments}
The work is supported by National Nature Science Foundation of China (No. 12164024 and No. 11864021), LanZhou Center for Theoretical Physics (No. 12247101).





\newpage

\clearpage
\begin{figure}[htbp] 
\centering 
\includegraphics[width=22cm]{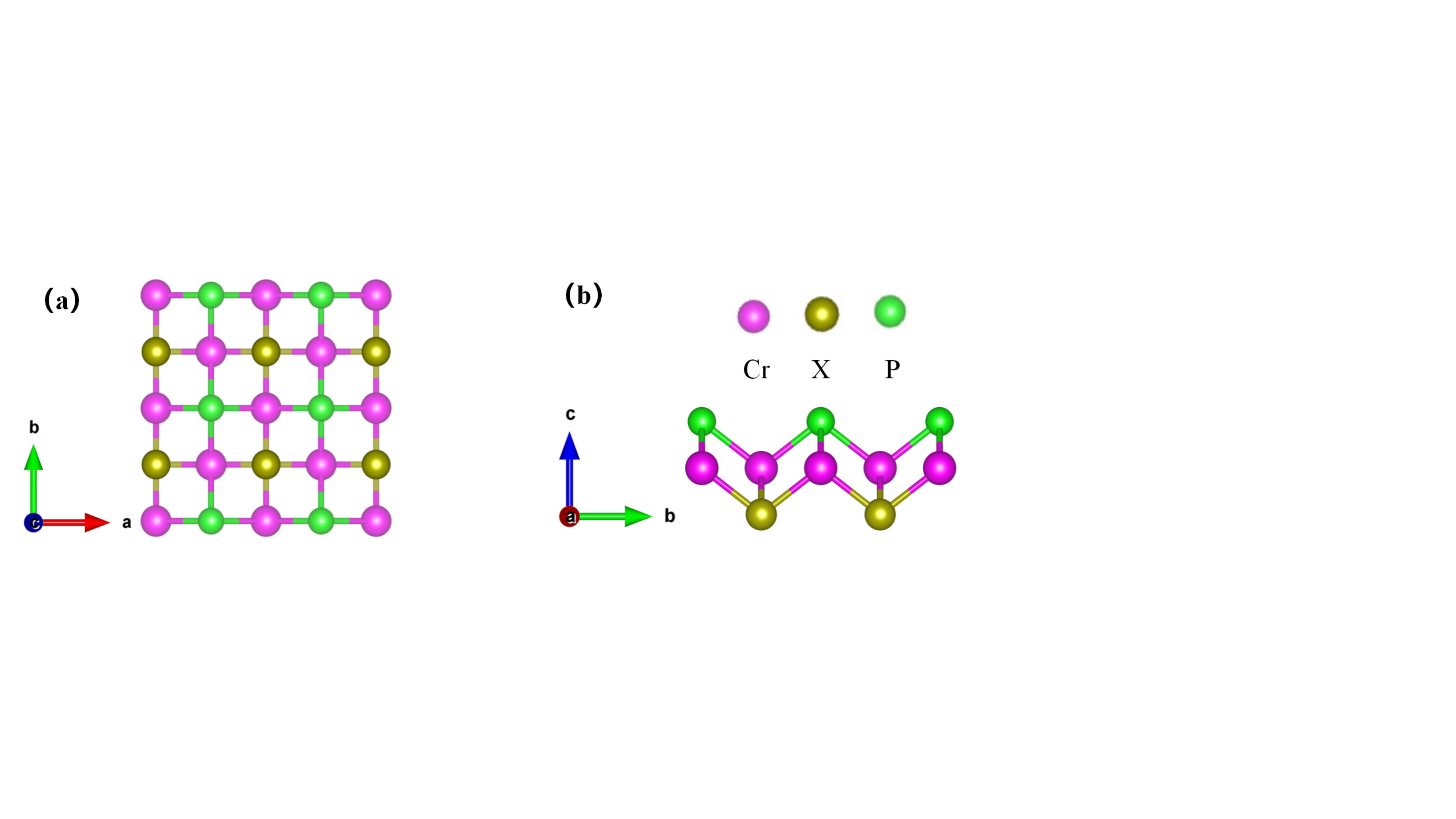}
\caption{(a) Top and (b) side views for two-dimensional material Cr$_{2}$XP (X=P, As, Sb).} 
\label{Fig1} 
\end{figure}

\clearpage
\begin{figure}[htbp] 
\centering 
\includegraphics[width=14.5cm]{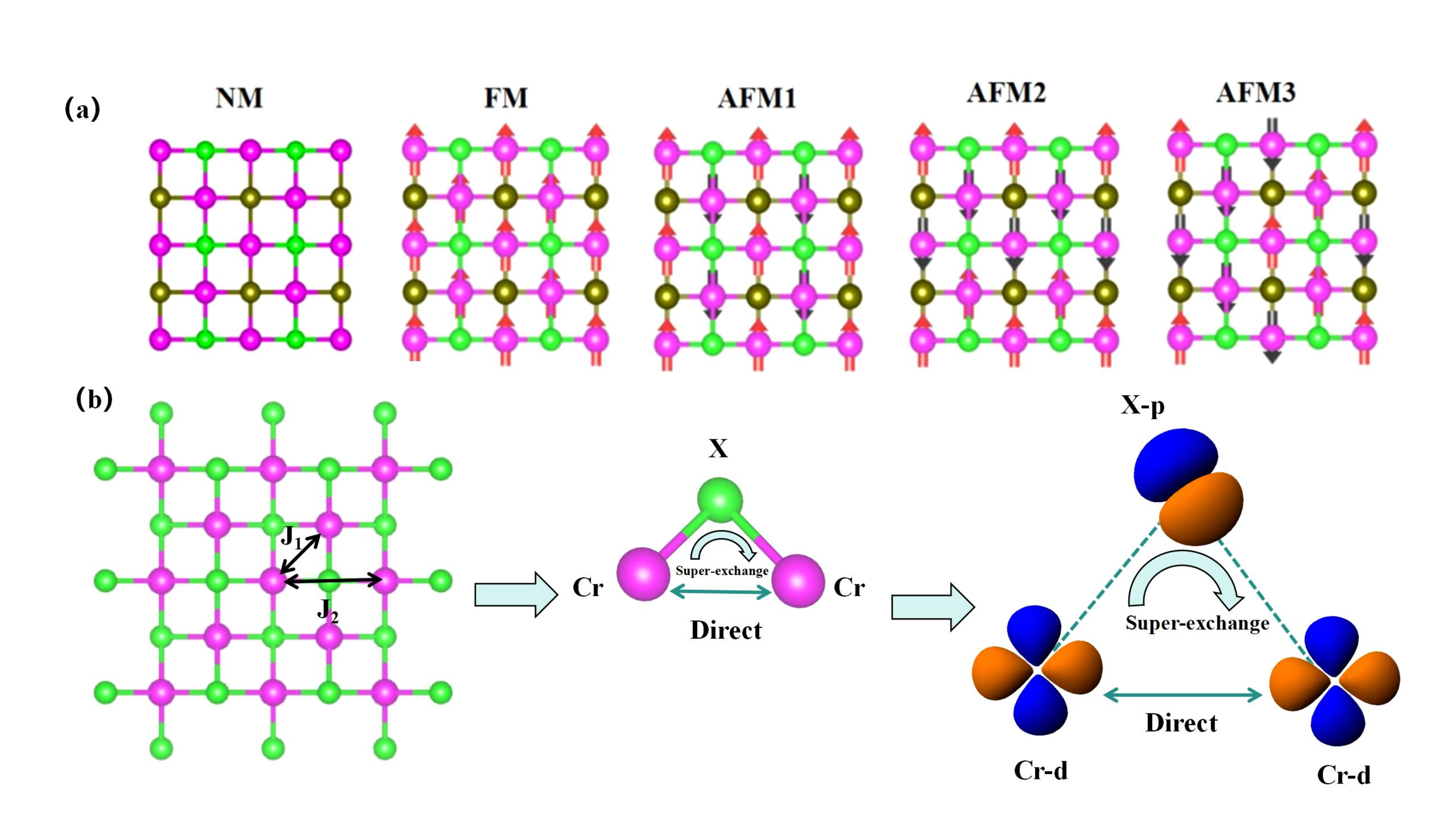} 
\caption{(a) The magnetic configurations of 2D material Cr$_{2}$XP (X=P, As, Sb). (b) The magnetic coupling mechanism for 2D material Cr$_{2}$XP (X=P, As, Sb) involves the nearest neighbor ($J_{1}$) and next-nearest neighbor ($J_{2}$) exchange interactions. } 
\label{Fig1} 
\end{figure}

\clearpage
\begin{figure}[htbp] 
\centering 
\includegraphics[width=14.14cm]{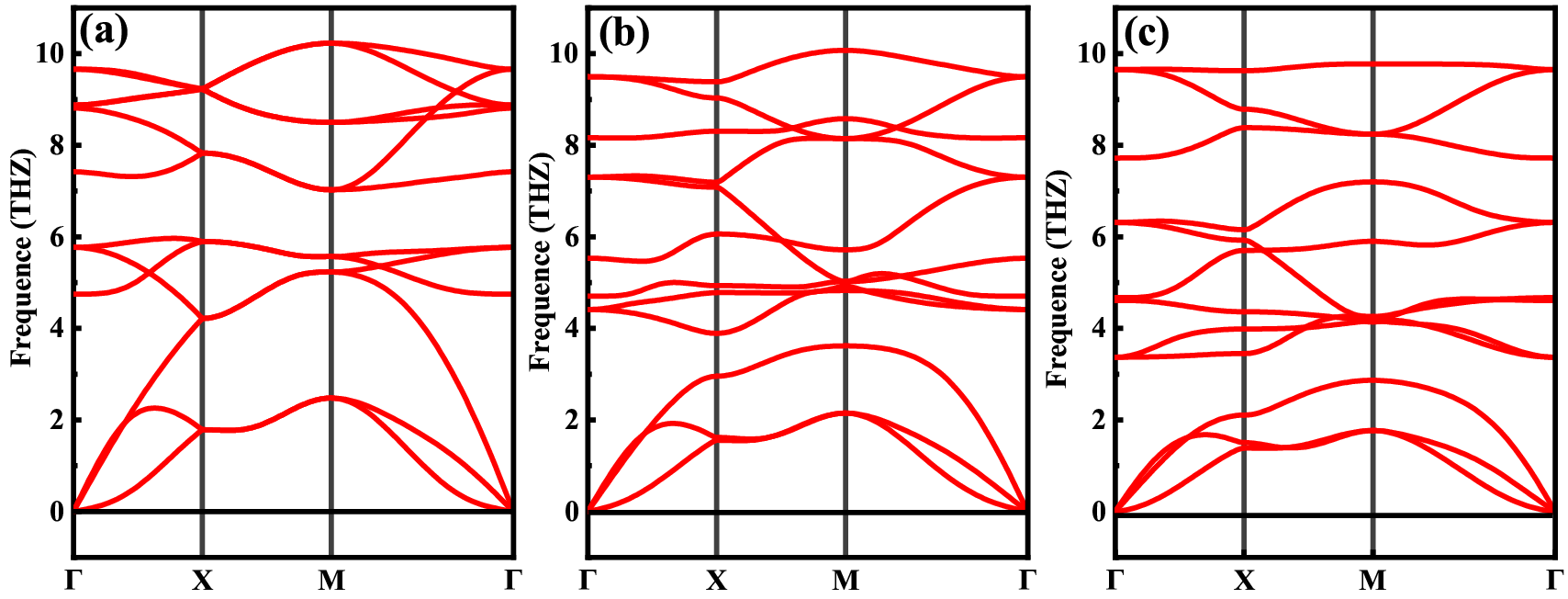} 
\caption{The phonon spectrums of (a) Cr$_{2}$P$_{2}$, (b) Cr$_{2}$AsP and (c) Cr$_{2}$SbP materials. } 
\label{Fig1} 
\end{figure}

\clearpage
\begin{figure}[htbp] 
\centering 
\includegraphics[width=14.5cm]{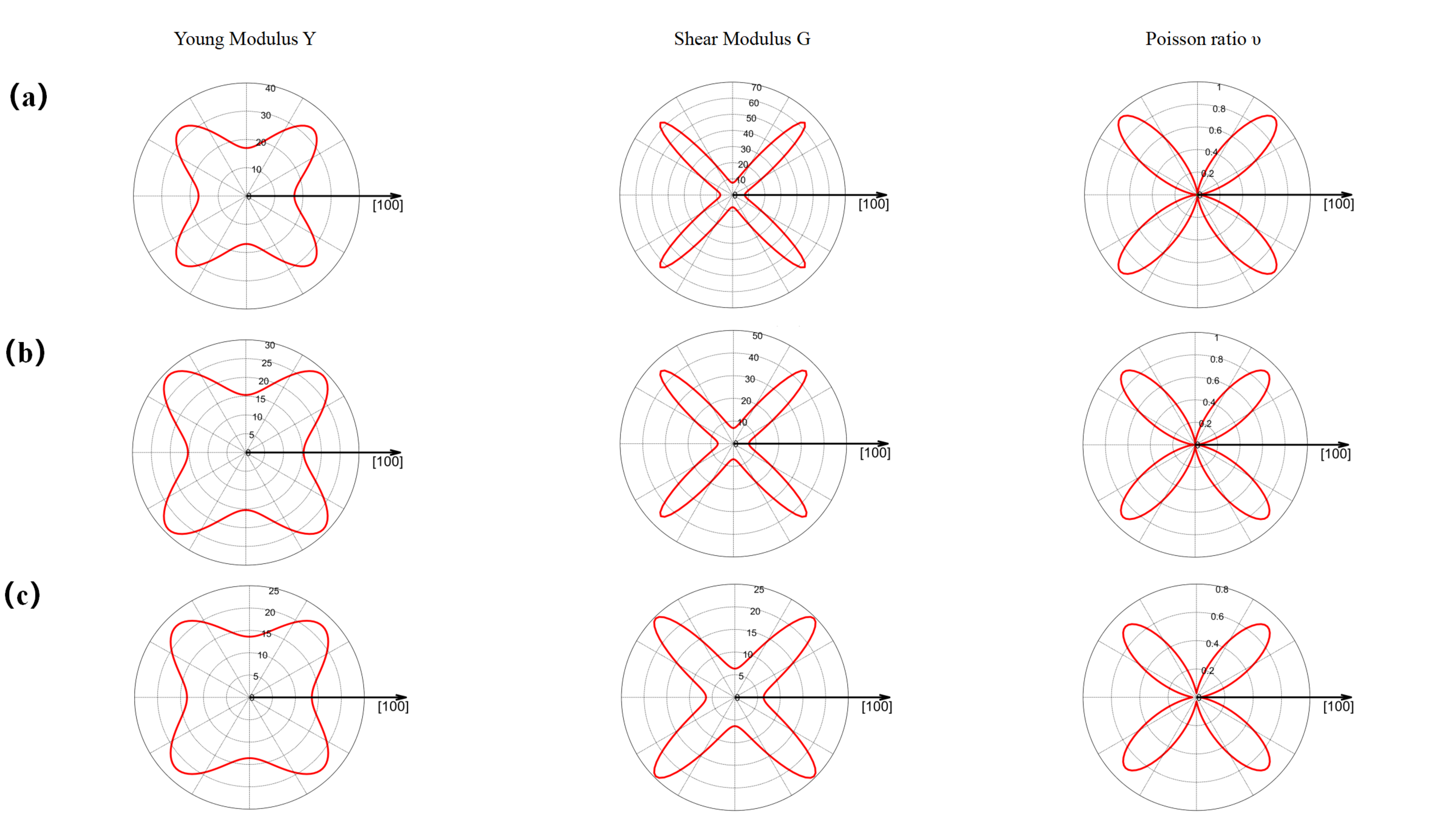} 
\caption{Two-dimensional elastic modulus of (a) Cr$_{2}$P$_{2}$, (b) Cr$_{2}$AsP and (c) Cr$_{2}$SbP materials. } 
\label{Fig1} 
\end{figure}

\clearpage
\begin{figure}[htbp] 
\centering 
\includegraphics[width=14.5cm]{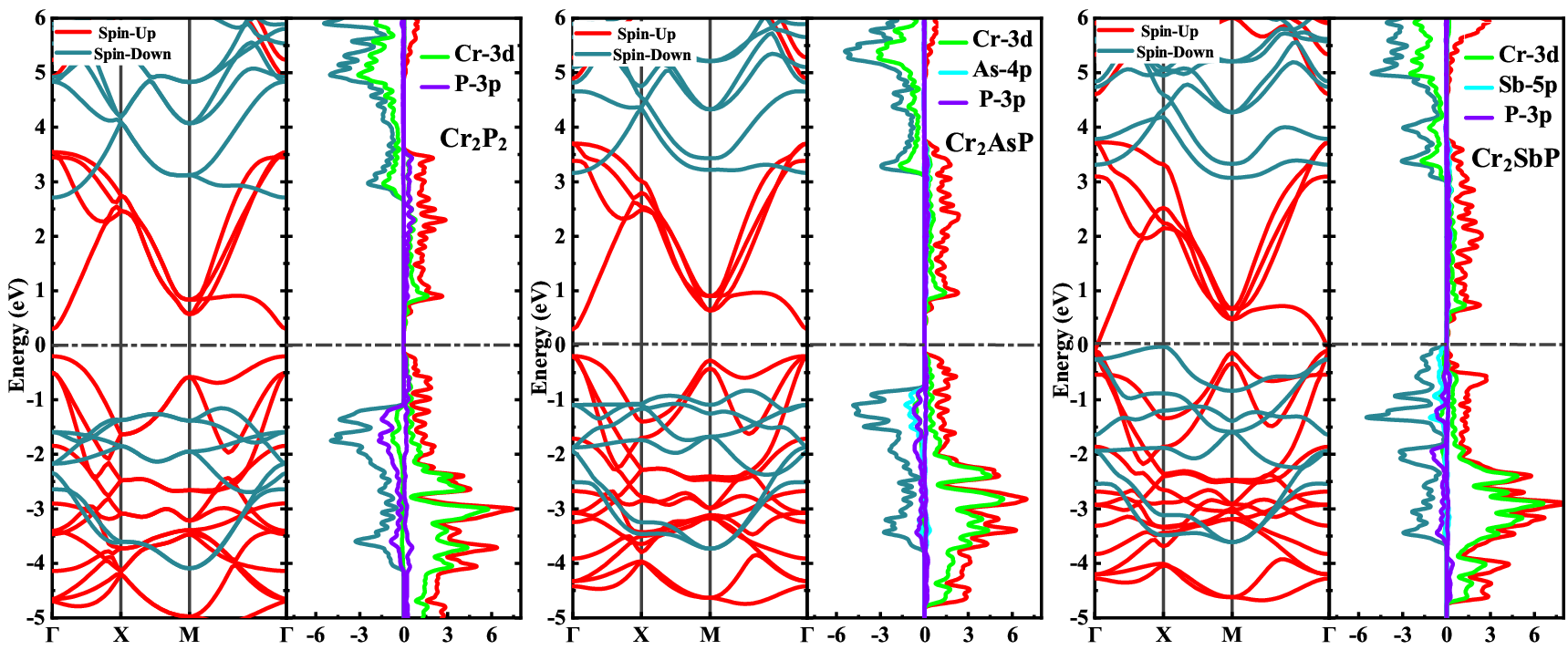} 
\caption{Band structure and density of states for two-dimensional material Cr$_{2}$XP (X=P, As, Sb).} 
\label{Fig1} 
\end{figure}

\clearpage
\begin{figure}[htbp] 
\centering 
\includegraphics[width=14.14cm]{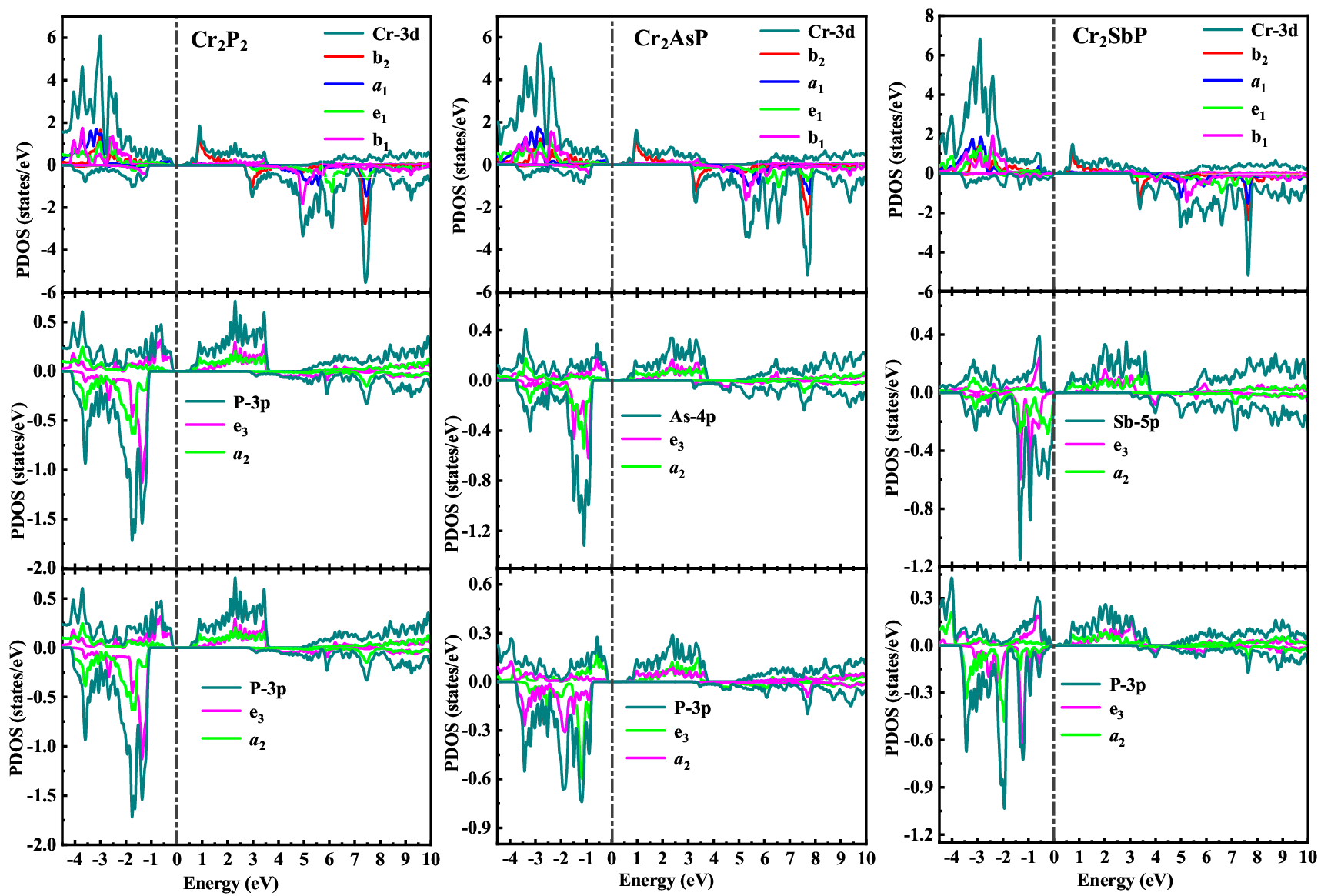} 
\caption{Local density of states for two-dimensional material Cr$_{2}$XP (X=P, As, Sb).} 
\label{Fig1} 
\end{figure}

\clearpage
\begin{figure}[htbp] 
\centering 
\includegraphics[width=14.14cm]{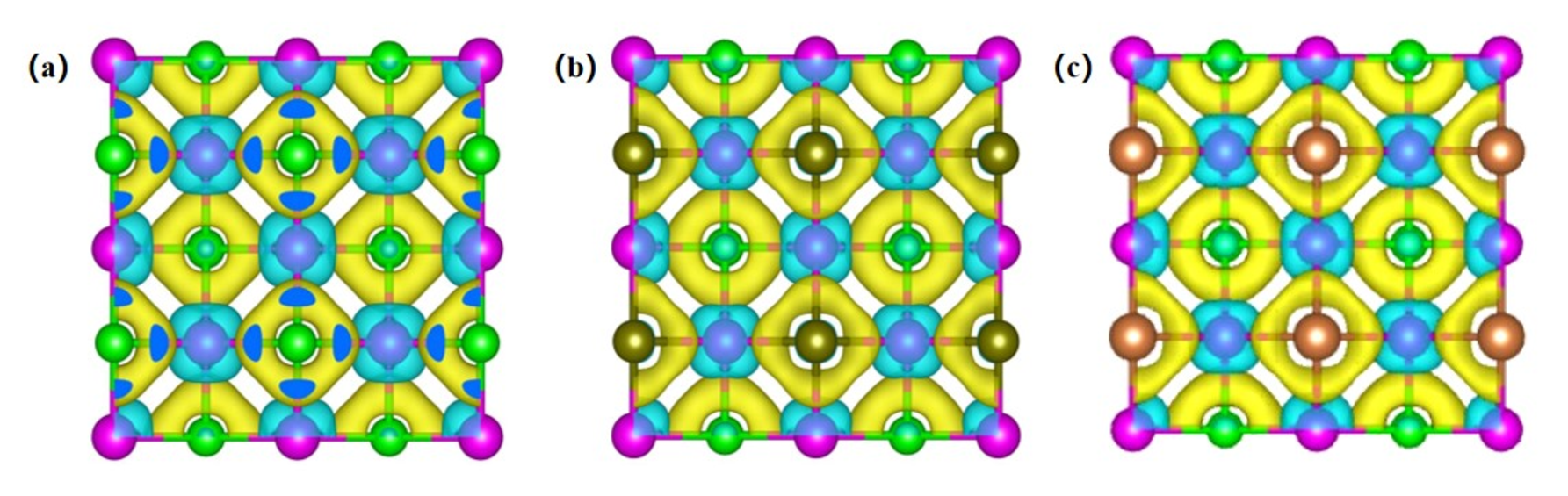} 
\caption{Differential charge densities for (a) Cr$_{2}$P$_{2}$, (b) Cr$_{2}$AsP and (c) Cr$_{2}$SbP.} 
\label{Fig1} 
\end{figure}

\clearpage
\begin{figure}[htbp] 
\centering 
\includegraphics[width=14.14cm]{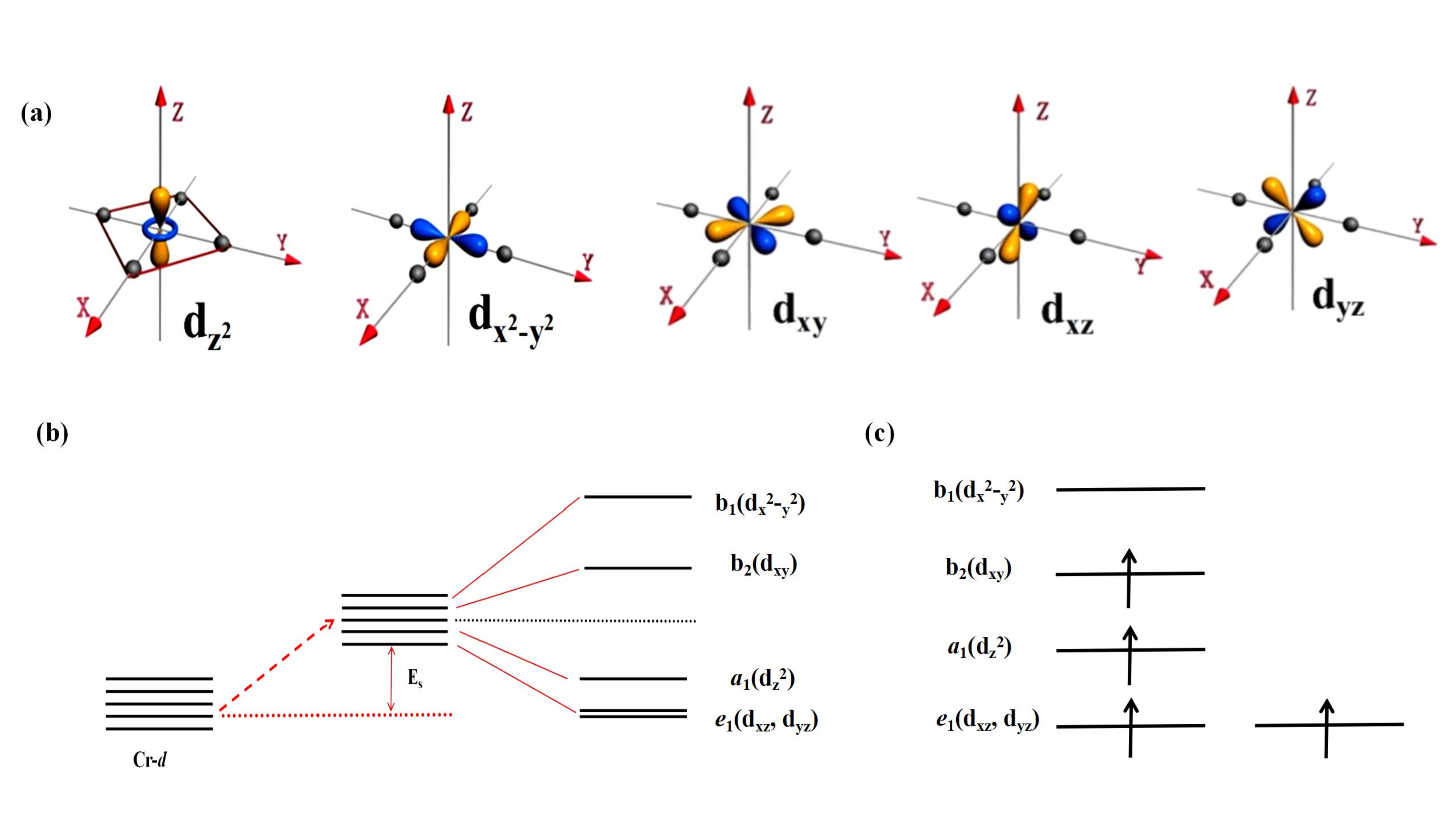} 
\caption{Electron Localization Functions (ELF) maps for (a) Cr$_{2}$P$_{2}$, (b) Cr$_{2}$AsP and (c) Cr$_{2}$SbP materials. (d, e, f) ELF line profiles along color lines in panel a-c.} 
\label{Fig1} 
\end{figure}

\clearpage
\begin{figure}[htbp] 
\centering 
\includegraphics[width=14.14cm]{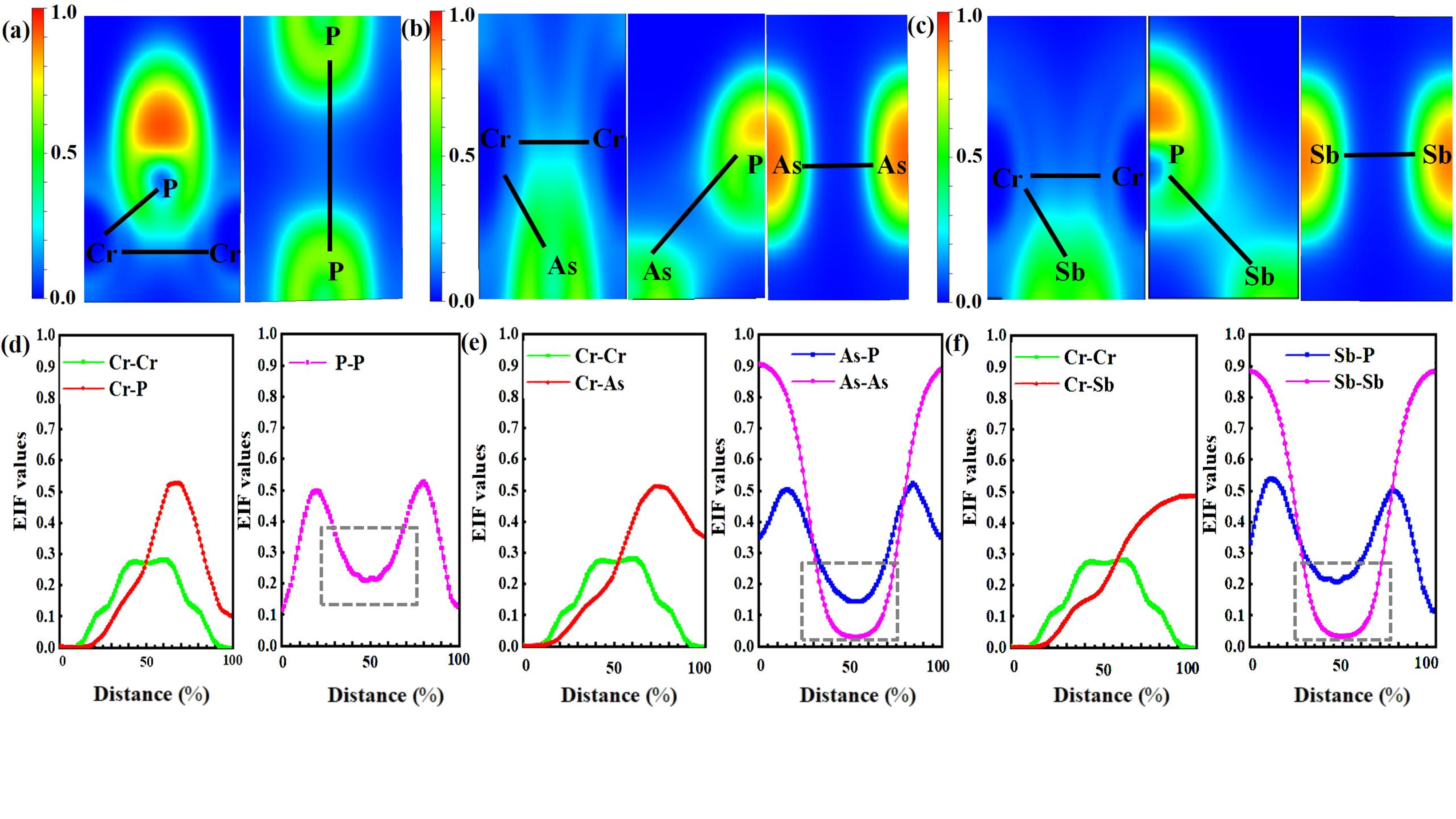} 
\caption{The splitting and occupation of Cr-$\textit{d}$ orbital energy level in a 2D square lattice field. } 
\label{Fig1} 
\end{figure}

\clearpage
\begin{figure}[htbp] 
\centering 
\includegraphics[width=14.14cm]{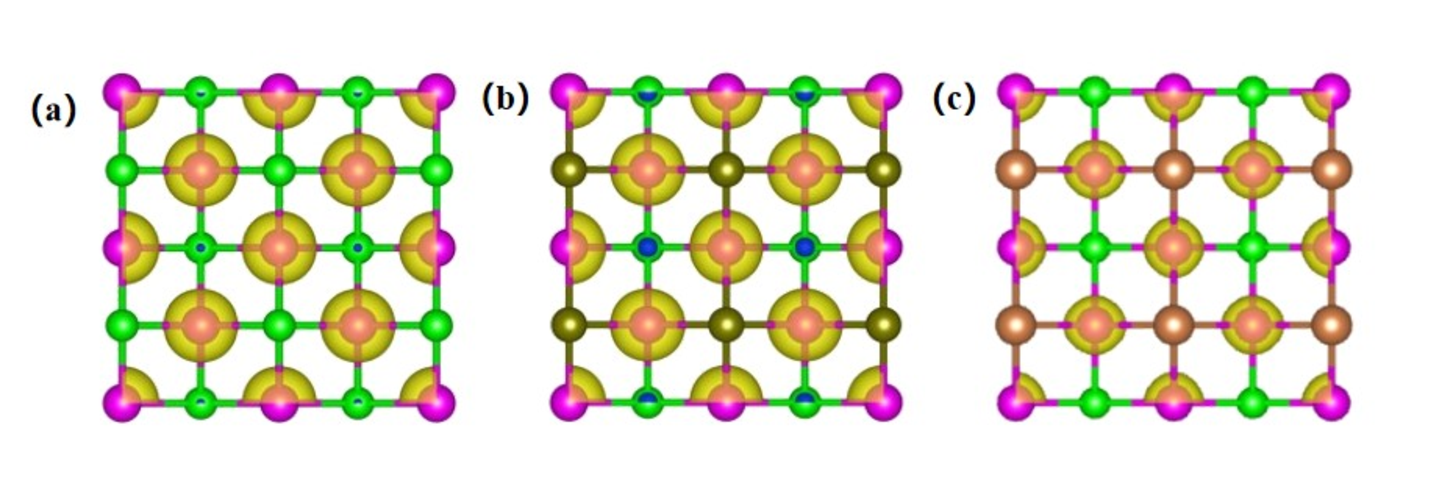} 
\caption{The spin charge densities for (a) Cr$_{2}$P$_{2}$, (b) Cr$_{2}$AsP and (c) Cr$_{2}$SbP.} 
\label{Fig1} 
\end{figure}


\clearpage
\begin{figure}[htbp] 
\centering 
\includegraphics[width=14.14cm]{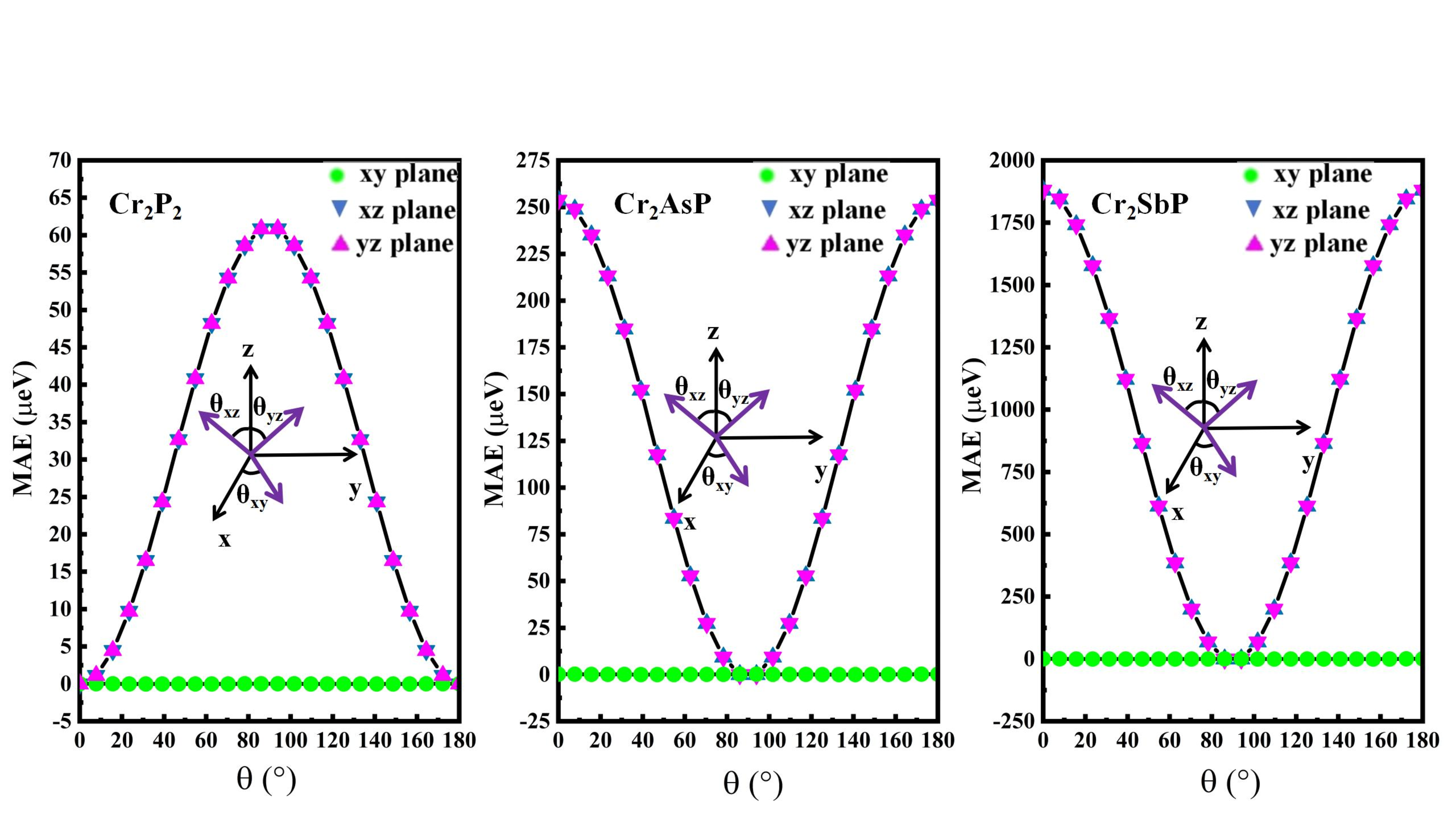} 
\caption{The magnetic anisotropy energy (MAE) for monolayer Cr$_{2}$XP (X=P, As, Sb) varies with the angle dependence of magnetization direction in three different planes. } 
\label{Fig1} 
\end{figure}

\clearpage
\begin{figure}[htbp] 
\centering 
\includegraphics[width=14.14cm]{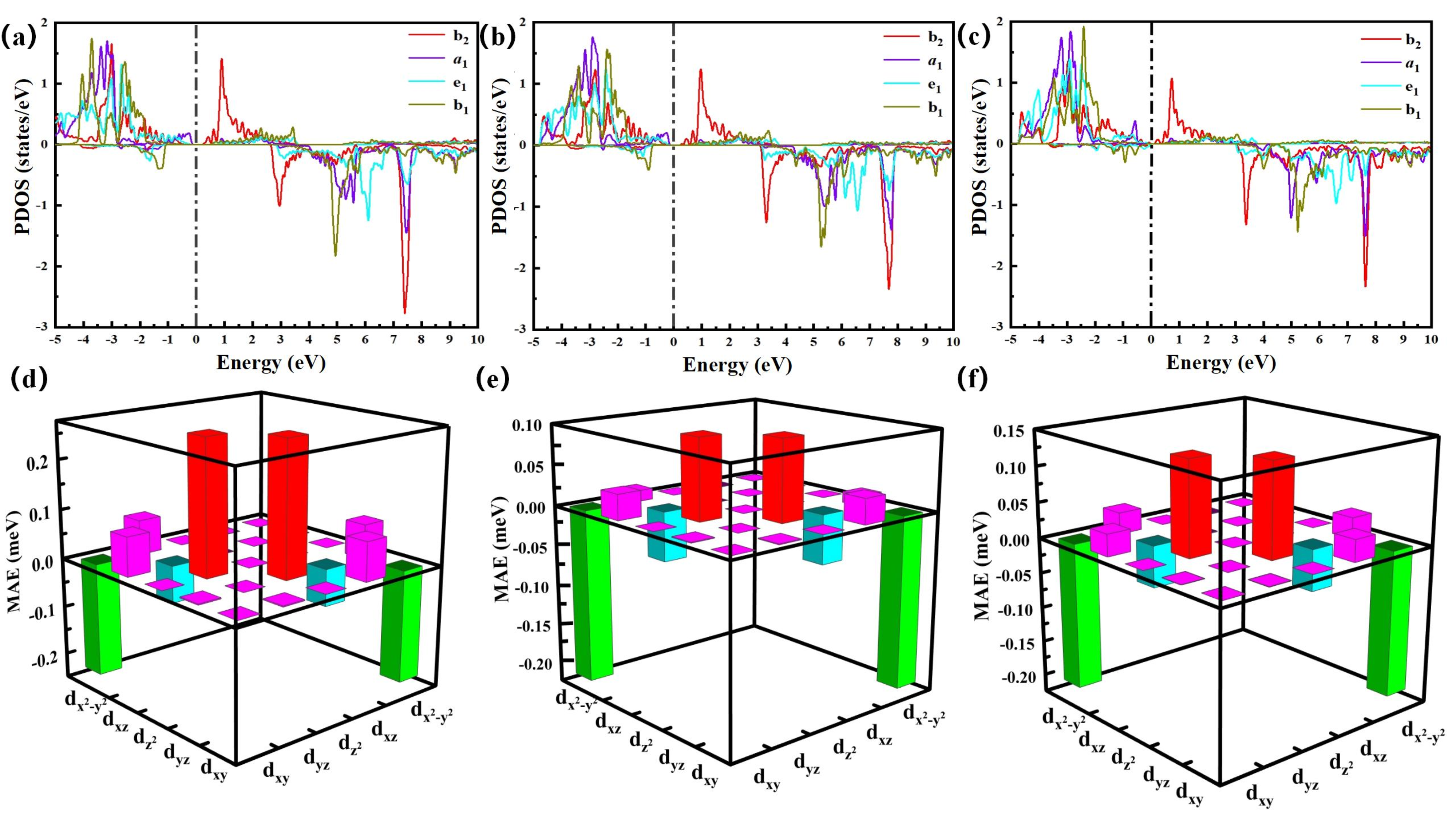} 
\caption{The Cr-3$d$-oribit-resolved MAEs and PDOS for two-dimensional material Cr$_{2}$XP (X=P, As, Sb).} 
\label{Fig1} 
\end{figure}

\clearpage
\begin{figure}[htbp] 
\centering 
\includegraphics[width=14.14cm]{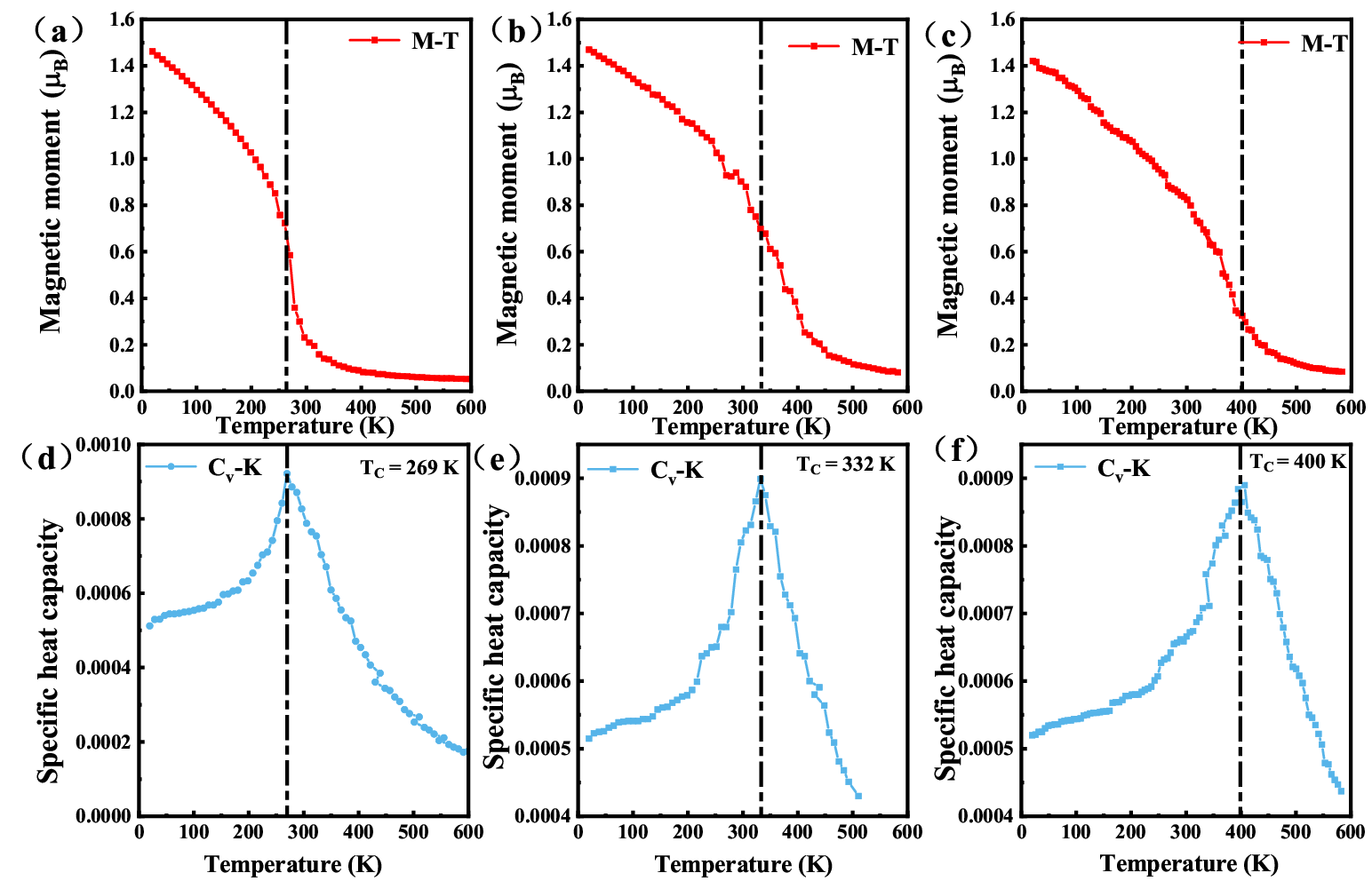} 
\caption{Magnetization for (a) Cr$_{2}$P$_{2}$, (b) Cr$_{2}$AsP and (c) Cr$_{2}$SbP with temperature. Specific heat capacity (C${_v}$) for (d) Cr$_{2}$P$_{2}$, (e) Cr$_{2}$AsP and (f) Cr$_{2}$SbP with temperature.} 
\label{Fig1} 
\end{figure}

\clearpage
\begin{figure}[htbp] 
\centering 
\includegraphics[width=14.14cm]{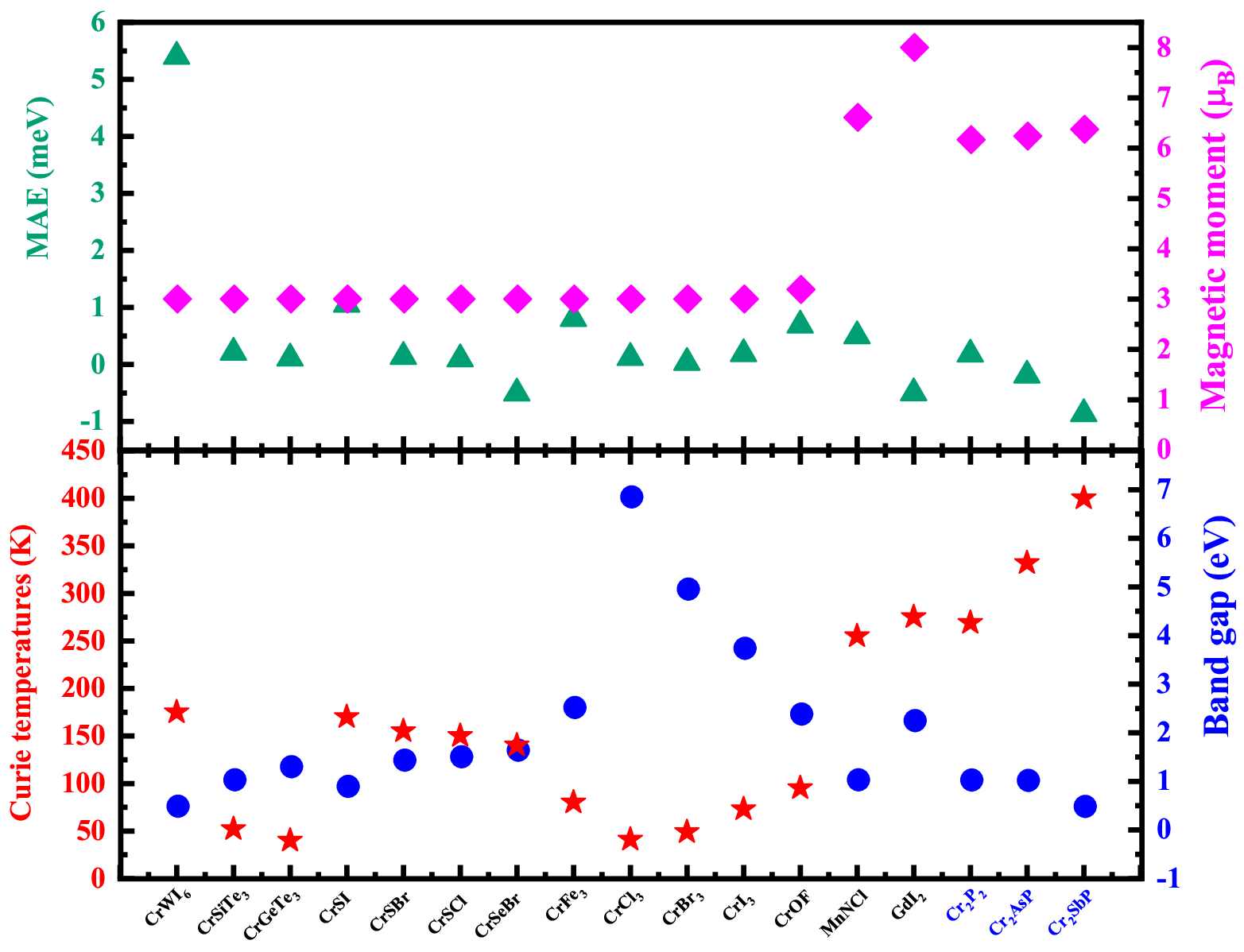} 
\caption{Summary of the magnetic anisotropy, total magnetic moment, band gap and Curie temperature for two-dimensional ferromagnetic semiconductors.} 
\label{Fig1} 
\end{figure}

\clearpage
\begin{table}[h]\footnotesize
\centering
\caption{\label{tab:test}
 Total energies (eV) for nonmagnetic (NM), ferromagnetic (FM) and anti-ferromagnetic (AFM) configurations for two-dimensional material Cr$_{2}$XP (X=P, As, Sb). }

\begin{tabular}{p{1.9cm}<{\centering} p{1.9cm}<{\centering} p{1.9cm}<{\centering} p{1.9cm}<{\centering} p{1.9cm}<{\centering} p{1.9cm}<{\centering}}
\hline
  Systems&NM&FM&AFM1&AFM2&AFM3\\
   \hline
    Cr$_{2}$P$_{2}$&-21.188938 &-25.303569 &-24.589492 &-24.787902 &-24.577904   \\
    Cr$_{2}$AsP    &-20.023325 &-24.483553 &-23.779039 &-23.962851 &-23.748328    \\
    Cr$_{2}$SbP    &-18.844686 &-23.526116 &-22.993635 &-23.098909 &-22.932443    \\
    \hline
 \end{tabular}
\end{table}

\clearpage
\begin{table}[h]\footnotesize
\caption{\label{tab:test}
The calculated lattice parameter $\textit{a}$$_{cal}$ ({\AA}), formation energy $E_{f}$ (eV/atom), elastic constants $C_{ij}$ (N/m) for two-dimensional material Cr$_{2}$XP (X=P, As, Sb). }

\begin{tabular}{p{1.5cm}<{\centering} p{1.5cm}<{\centering} p{1.5cm}<{\centering} p{1.5cm}<{\centering} p{1.5cm}<{\centering}p{1.5cm}<{\centering}p{1.5cm}<{\centering}}
\hline
  Systems&$\textit{a}$$_{cal}$&$E_{f}$&$C_{11}$&$C_{12}$&$C_{22}$&$C_{66}$\\
   \hline
    Cr$_{2}$P$_{2}$&4.17 &-0.94 &36.76 &17.99 &36.76 &25.14 \\
    Cr$_{2}$AsP    &4.23 &-1.08 &33.68 &16.30 &33.68 &21.30  \\
    Cr$_{2}$SbP    &4.29 &-1.11 &29.18 &12.44 &29.18 &16.12  \\
    \hline
 \end{tabular}
\end{table}

\clearpage
\begin{table}[h]\footnotesize
\caption{\label{tab:test}
The calculated Young's modulus $Y_{s}^{2D}$ (N/m), shear modulus \textit{G} (N/m), Poisson's ratio $\nu$, thickness of the nano-sheet $\textit{l}$ ($\mu$m), density $\sigma$ (Kg/m$^{2}$), gravity deformation $h/\textit{l}$ for two-dimensional material Cr$_{2}$XP (X=P, As, Sb). }

\begin{tabular}{p{1.3cm}<{\centering} p{1.3cm}<{\centering} p{1.3cm}<{\centering} p{1.3cm}<{\centering} p{1.3cm}<{\centering} p{2.5cm}<{\centering}p{1.9cm}<{\centering}}
\hline
  Systems&$Y_{s}^{2D}$&$\textit{G}$&$\upsilon$&$\textit{l}$&$\sigma$ &$h/\textit{l}$\\
   \hline
    Cr$_{2}$P$_{2}$&27.96 &25.14 &0.49 &100 &0.962$\times$10$^{-6}$   &$3.230\times$10$^{-4}$  \\
    Cr$_{2}$AsP    &25.79 &21.30 &0.48 &100 &1.161$\times$10$^{-6}$   &$3.530\times$10$^{-4}$  \\
    Cr$_{2}$SbP    &23.87 &16.12 &0.43 &100 &1.319$\times$10$^{-6}$   &$3.780\times$10$^{-4}$  \\
    \hline
 \end{tabular}
\end{table}

\clearpage
\begin{table}[h]\footnotesize
\caption{\label{tab:test}
Bader charge for Cr$_{2}$XP (X=P, As, Sb) material. Note: ``+'' means losing electrons, ``-'' means gaining electrons. }

\begin{tabular}{p{3cm}<{\centering} p{3cm}<{\centering} p{3cm}<{\centering} p{3cm}<{\centering}}
\hline
  Systems&Cr&X&P\\
   \hline
    Cr$_{2}$P$_{2}$&+1.20 &-1.20 &-1.20   \\
    Cr$_{2}$AsP    &+1.11 &-0.97 &-1.25   \\
    Cr$_{2}$SbP    &+0.95 &-0.63 &-1.29  \\
    \hline
 \end{tabular}
\end{table}

\clearpage
\begin{table}[h]\footnotesize
\setlength{\tabcolsep}{3pt}
\caption{\label{tab:test}
The total $M_{t}$ and atomic magnetic moments ($\mu$$_{B}$) for two-dimensional material Cr$_{2}$XP (X=P, As, Sb).}
\begin{tabular}{p{2.5cm}<{\centering} p{2.5cm}<{\centering} p{2.5cm}<{\centering} p{2.5cm}<{\centering}p{2.5cm}<{\centering}}
\hline
  Systems&Cr&X&P&$M_{t}$\\
   \hline
    Cr$_{2}$P$_{2}$&3.439 &-0.357&-0.357 &6.165 \\
    Cr$_{2}$AsP    &3.511 &-0.411&-0.366 &6.237   \\
    Cr$_{2}$SbP    &3.603 &-0.423&-0.373 &6.372  \\
    \hline
 \end{tabular}
\end{table}

\clearpage
\begin{table}[h]\footnotesize
\setlength{\tabcolsep}{2pt}
\caption{\label{tab:test}
Total energy (eV) for different magnetization directions, MAE ($\mu$eV) and $T_{C}$ (K) for two-dimensional material Cr$_{2}$XP (X=P, As, Sb).}

\begin{tabular}{p{2.1cm}<{\centering} p{2.1cm}<{\centering} p{2.1cm}<{\centering} p{2.1cm}<{\centering}p{2.1cm}<{\centering}p{2.1cm}<{\centering}}
\hline
  Systems&[100]&[010]&[001]&MAE &$T_{C}$\\
   \hline
    Cr$_{2}$P$_{2}$&-25.31085880  &-25.31085880 &-25.31091977  &60.97  &269 \\
    Cr$_{2}$AsP    &-24.50360720  &-24.50360720 &-24.50334909  &-258.11 &332   \\
    Cr$_{2}$SbP    &-23.60499465  &-23.60499465 &-23.60308968  &-1904.97 &400  \\
    \hline
 \end{tabular}
\end{table}

\end{document}